\documentstyle[sprocl,epsfig]{article}

\def\be{\begin{equation}}
\def\ee{\end{equation}}
\def\bea{\begin{eqnarray}}
\def\eea{\end{eqnarray}}

\begin{document}

\title{Photons from the Universe\\
New Frontiers in Astronomy}
\author{Hinrich Meyer}

\address{University of Wuppertal, Gau\ss stra\ss e 20,\\ 42097 Wuppertal,
Germany}

\maketitle\abstracts
{This century has seen a dramatic increase in observational
possibilities of the physics of the universe. Several of the very recent new
developments with emphasis on the particle physics aspect and in particular $%
\gamma\gamma$ interactions are briefly discussed in this talk.}

\section{The universal photon background}

\label{The universal photon background}

\begin{figure}[tb]
\begin{center}
\psfig{figure=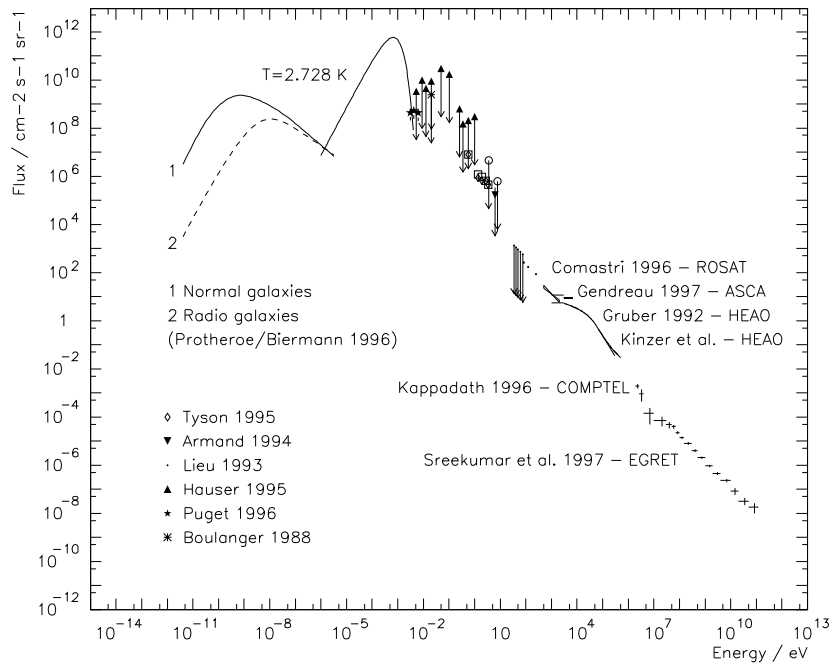,width=11cm}
\end{center}
\caption{Flux of the diffuse extragalactic background radiation.}
\label{gups}
\end{figure}

\begin{figure}[tb]
\begin{center}
\psfig{figure=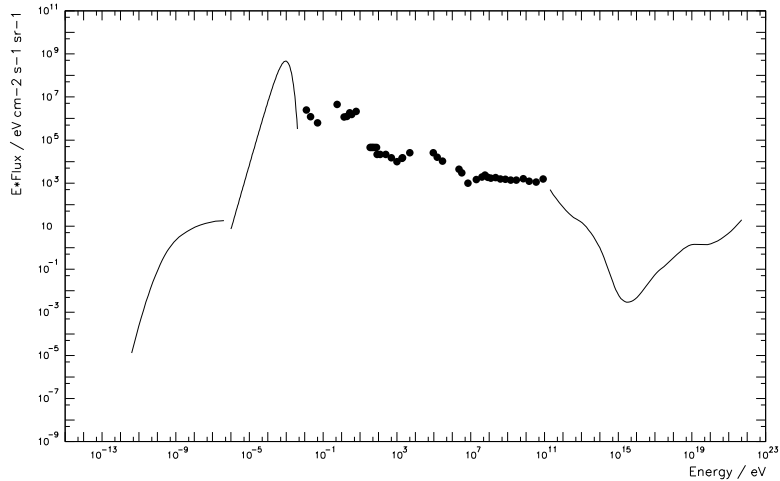,width=11.0cm}
\end{center}
\caption{The energy flux of the intergalactic universal photon background.
The line above 100 GeV indicates a model calculation only, since no
measurements are available yet in this energy range.}
\label{plotmalenergie}
\end{figure}

At any point between galaxies in the universe one encounters a field of
photons ranging in energy from the long wave end of the radiospectrum at $%
10^{-12}eV$ to at least $10^{11}eV$ and possibly $10^{20}eV$ which is the
upper end of the cosmic ray spectrum observed at earth. To a large extend
astronomy is based on the observation of local deviations from this
universal background and was based over millenia on photons in a very narrow
energy interval at about $1eV$ accessible for detection by the human eye.
Since the 1940th the radio range was explored and after 1960 the X-ray, the
MeV and GeV range, and the infrared and extreme ultraviolet mostly using
instruments launched by rockets and operating outside the earth atmosphere.
The photon energy range to be explored with these new experimental
possibilities expanded by many orders of magnitude and entirely new
processes in the universe became available as tools of astronomy. As one
result of the observations it now becomes possible to construct the
universal photon flux for all wavelengths up to about 100 GeV photon energy.
To achieve this one has to eliminate the local foreground flux of photons
with the galaxy obviously as the most prominent local enhancement visible
over the whole energy spectrum. In the infrared region only upper limits are
available, due to the brilliance of our very local environment in this
energy range. The universal photon background was given about 10 years ago
by Ressell and Turner~\cite{res}. Since then, new information became
available, at almost all wavelengths and an attempt has been made to update
the compilation in~\cite{hoh}. This is shown in figure \ref{gups} as photon
flux in units of $cm^{-2}s^{-1}sr^{-1}$ and figure \ref{plotmalenergie} as
energy flux per unit decade in eV. The highest flux originates from rather
early in the universe at a black body temperature of presently $\approx
2.7^{\circ }K$. It is a pure Planck spectrum exact to a very high degree,
with dipole distortion due to the motion of the solar system and remaining
nonuniformities at the $10^{-5}$ level indicating seeds of structure
formation in an expanding universe~\cite{fix}.

In the radio range the source of photons is synchrotron radiation of
relativistic electrons in ambient magnetic fields mostly from normal radio
galaxies with a sizeable contribution from a small subclass of {\bf g}%
alaxies with a highly {\bf a}ctive {\bf n}ucleus (AGN)~\cite{pro}. The AGNs
are also contributing a dominant part of the X-ray region~\cite{set} as well
as in the MeV - GeV range, which has only very recently been revealed from
observations using the Compton Gamma Ray Observatory (CGRO)~\cite{sre}.
Beyond 100 GeV the photon background is not yet known; shown here is a
theoretical estimate~\cite{lee} based on a universal flux of cosmic rays at
energies $>10^{19}eV$ that interact and cascade in the universal photon
background. This cascade starts with single pion photoproduction of protons
on photons from the $2.7^{\circ }K$ field~\cite{gre} and develops with short
cascade length until all photons have energies $\ll 100\;TeV$ where the
interaction length becomes very large again~\cite{wdo}. The energy range
between 100 GeV and 100 TeV is at present the frontier region of astronomy
with photons ( TeV - astronomy ). For energies $>$ 100 TeV the density of
the photon background does not allow to look beyond our own galaxy (see
figure \ref{horizon}). Here neutrinos may step in and at energies $>10^{19}eV
$ even protons depending on the structure and strength of magnetic fields in
and beyond our local cluster. It thus appears that the energy range $>$100
GeV is a true domain of high energy physics in the universe where processes
usually studied in the laboratory at particle accelerators are used to
reveal the nature of extreme objects and environments in space, with the
very early universe at temperatures $>$100 GeV as the most prominent and
singular spot in spacetime.

\section{The process $\gamma \gamma \rightarrow e^{+} e^{-}$ in the universe}

\label{The process gamma gamma}

Soon after the discovery of the positron the cross--section for the
fundamental process photon + photon $\rightarrow$ electron + positron was
calculated~\cite{jau}. For an incoming photon of energy E colliding with a
target photon of energy $\varepsilon$ at an angle $\Theta$ the threshold
energy E$_{th}$ is given by 
\begin{eqnarray}
E_{th} & = & \frac{ 2 \cdot m_{e}^{2} }{\varepsilon (1-\cos\Theta)}  \nonumber
\end{eqnarray}
and in units relevant for TeV astronomy: 
\begin{eqnarray}
E_{th} & \simeq & \frac{1}{2} \left[ \frac{1 \: TeV}{\varepsilon} \right] eV
\end{eqnarray}
\label{eq:threshold} The total cross--section for the process is given as~%
\cite{gou}

\begin{center}
\begin{eqnarray}
\sigma_{\gamma\gamma} = \frac{3}{16} \sigma_{0}(1 - \beta^{2}) \left[ (3 -
\beta^{4}) \, \ln\, \frac{1 + \beta}{1 - \beta} - 2\beta (2 - \beta^{2}) \right]
\end{eqnarray}
\label{eq:crosssection}
\end{center}
with
\begin{center}
\[
\beta \equiv \left( \frac{1 - 2 m_{e}^{2} }{E \varepsilon (1 - \cos\Theta) }
\right)^{\frac{1}{2}} 
\]
\label{eq:beta}
\end{center}
and
\begin{center}
\[
\sigma_{0} = 6.65 \, \cdot \, 10^{-25} \; cm^{2} 
\]
\label{eq:sigmao}
\end{center}
the Thompson cross section.

\begin{figure}[th]
\begin{center}
\epsfig{figure=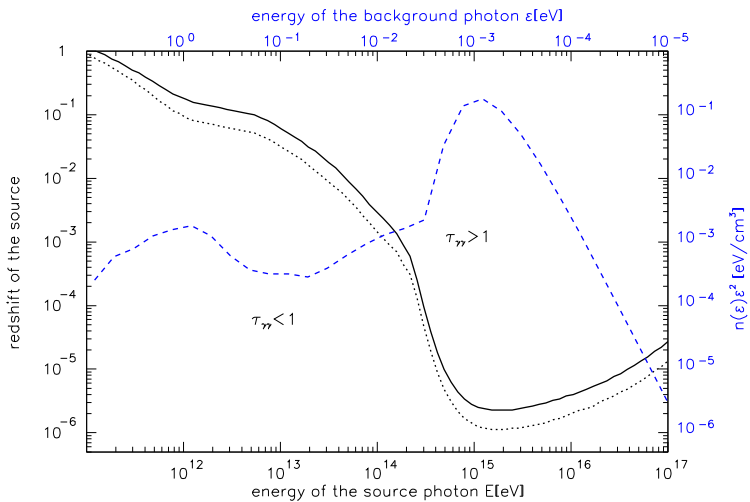,width=10.0cm}
\end{center}
\caption{The $\gamma $--ray horizon for photons of energy $>10^{11}\,eV$
(solid line). It is based on a photon background density as shown by the
dashed line (and right hand scale). The region of optical and infrared
photons is described here by a model calculation$^{12}$}
\label{horizon}
\end{figure}

The cross section rises rapidly after threshold with a peak value of about
200 mbarn at about 2$\cdot E_{th}$ and then it falls off approximately as
1/E. This behavior of the cross section folded with the $2.7^{\circ }K$
Planck spectrum then results in the deep absorption trough at about 2 PeV
shown in figure \ref{horizon} taken from \cite{fun}. At about 1 eV light from
stars dominates and absorbs most strongly at about 1 TeV while the far
infrared part of the $2.7^{\circ }K$ photons should cut off all photons at
about 150 TeV. The transparency in the window (1 - 150) TeV is very
uncertain, as the universal photon flux in the range from 6$\cdot $10$^{-3}$
eV to 6$\cdot $10$^{-1}$ eV is not measured yet and can be estimated only
from rather involved models of dust- and star formation throughout the
lifetime of the universe. Several such calculations have become available
recently and can be used to calculate a Gamma Ray Horizon see figure \ref
{horizon} and \cite{ste}. 
It is obvious that considerable uncertainty is present on how deep into
space high energy photon sources can be revealed. Interesting structures in
the universe are nearby galaxies at say 1 Mpc that could not be seen above
100 TeV while below 100 TeV the local cluster and beyond is probably
accessible, although the uncertainty on how far one will be able to see is
presently very much open. One of the very challenging tasks of TeV astronomy
is the exploration of the $\gamma $-ray horizon and various ideas on
possible sources at the horizon have been published recently~\cite{man}.

\section{The pair--compton cascade}

\label{The pair compton cascade} As can be seen from figure \ref{horizon}
the absorption of $>$100 TeV photons by the ubiquitous $2.7^{\circ }K$
photon background has a characteristic length of $\sim$10 kpc, which is
short compared to intergalactic distances. The resulting pair electrons have
a similarity short Compton scattering length and therefore an intergalactic
pair--Compton cascade develops as long as there are photons left in the
cascade with energy $\gg$100 TeV. The $e^{+}e^{-}$ pairs 'see' intergalactic
magnetic fields, or stronger fields near galaxies which results in
measurable delay phenomena of bursts~\cite{pla}, the creation of so called
pair halos near the source~\cite{aha}, or changes in the power law spectra
due to pile up of cascade photons~\cite{pro2}. This will serve as great
tools to explore intergalactic magnetic fields at values $< 10^{-11}G$ that
are hardly accessible by other experimental means.

\section{Active galactic nuclei (AGN)}

\label{AGN} Galaxies with an active nucleus have been known for long time
and have been the subject of countless investigations. A particularly
interesting feature is the high variability of the emission from the
nucleus. In the radio and the optical sizeable polarization has been
detected, which points to synchrotron radiation at the source. At radio
frequencies thanks to interferometry between antenna far apart even up to
intercontinental distances spatial resolutions at better than 1/1000 of an
arcsec can be obtained. These observations resolved the AGN emission into a
succession of 'blobs' seemingly moving with superluminal velocity (v$>$c)
across the sky. If the train of 'radioblobs' consists of relativistic
plasma (emitting the radio photons) with Lorentzfactors $\gg $1 moving
towards us at small angle $<10^{\circ }$ in a jet like fashion,
consideration of Lorentz transformation gives for the transverse velocity ($%
\beta $) of the blob 
\begin{center}
\begin{eqnarray}
\beta\;=\;\frac{\beta_{j}\sin\Theta}{1-\beta_{j}\cos\Theta}\; \simeq \;\frac{2%
}{\Theta} \:\:\;\;\; (for \;  \Theta \ll 1 \; and \; \beta_{j} \rightarrow 1)
\end{eqnarray}
\label{eq:velocity}
\end{center}
with $\beta_{j}$ the velocity of the blob and $\Theta$ the viewing angle.
This explains the apparent superluminal motions observed in many jets of
AGN. As a further consequence the observed luminosity $L_{obs}$ of the
source is considerably enhanced over the restframe luminosity $L$ according
to 
\begin{eqnarray}
L_{obs} & = & D^{3}_{j} L
\end{eqnarray}
\label{eq:luminosity}
\begin{center}
with 
\[
D_{j} = \frac{1}{\gamma_{j}(1-\beta_{j}\cos\Theta)} \;\;\; and \;\;\;
\gamma_{j} = \left( \frac{1}{ ( 1-\beta_{j}^{2})} \right)^{\frac{1}{2} } 
\]
\end{center}

The AGN should in fact not only have a jet pointing towards the observer but a
balancing jet receding in the opposite direction. The same relativistic
transformation effect responsible for the high brilliance of the jet towards
us renders the receding jets hardly observable. At larger inclination angle
however both jets become visible. Small changes in the viewing angle for the
jet moving towards us may be responsible for the large and somewhat
stochastic nature of the intensity variations as it may result from moving
plasma at relativistic velocity along helical paths. There is indeed
evidence for this from radio observations~\cite{kri}.

It is assumed that the jets originate from a central black hole of mass $%
10^{8}$-$10^{11}$ solar masses. The black hole accretes matter from a flat
disk spinning at relativistic speed at it's inner edge not far from the
horizon of the black hole. This setup may be surrounded by a huge dust torus
connecting up with stars and interstellar matter of the host galaxy.
Different viewing angles then produce a large variety of observational
phenomena for the AGNs.

The energy flux versus wavelength for AGN is rather flat within 1-2 orders
of magnitude from the radio range over more than 20 orders of magnitude into
the GeV range. It renders AGN as prominent contributors to the universal
photon flux in the universe at almost all energies (see figure \ref{gups}).
The most notable exception is of course the $2.7^{\circ}K$ microwave
background that originates from the big bang.

\section{AGN at GeV-Energies}

\label{AGN at Gev Energies} Launched in spring 1991 the $\underline{C}ompton$
$\underline{G}amma$ $\underline{R}ay$ $\underline{O}bservatory$ (CGRO) has
provided the first all sky survey in the high energy (up to 100 GeV) gamma
ray range using the EGRET instrument. As a result many new sources have been
discovered and in addition the diffuse gamma emission from the galaxy and
from extragalactic space have been determined rather accurately~\cite{str}.
Among the sources the identification of 65 AGN at the $>$(4-5) $\sigma$
level stands out as a great discovery. Several of these AGN have - when
flaring - the highest energy flux in the GeV region, they all show dramatic
variability in flux, and the spectra are rather flat with power law indices on
average about 2. They constitute a fraction $>$10$\%$ of all known flat
spectrum AGN, out to a distance of z=2.5~\cite{har}. The contrast in
amplitude is large about an order of magnitude, seemingly larger then at the
other wavelengths. Since EGRET provided only rather moderate sensitivity it
is not unreasonable to assume that mostly the peaks of the emission have
been seen and future missions with higher sensitivity and also better
exposure (e.g. GLAST) may detect all flat spectrum AGN's. These sources also
may provide a sample of candidates for ``Beacons at the Gamma Ray Horizons~%
\cite{man}'' if the energy spectra could be followed to higher energies. The
more distant ones should hit the horizon at ten's of GeV (see figure \ref
{horizon}). It is therefore of great importance to increase the sensitivity
of GeV-TeV gamma ray instruments to finally detect the absorption feature in
the energy spectra due to the universal photon background in the universe.

\section{The new frontier - TeV energies}

\label{The new frontier TeV energies}

\begin{figure}[tb]
\begin{center}
\psfig{figure=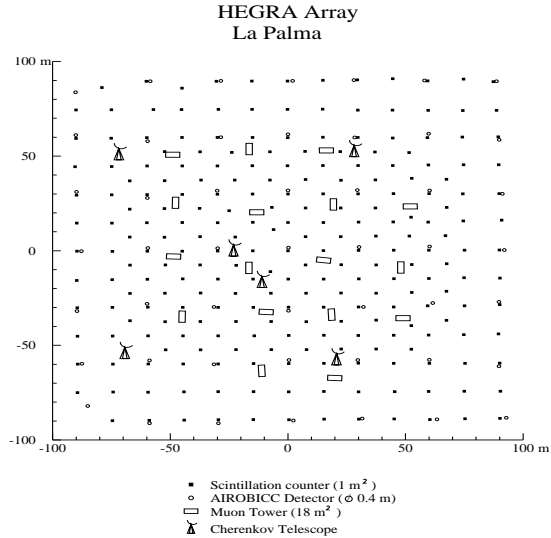,width=7.0cm,height=8.0cm}
\end{center}
\caption{Layout of the HEGRA experiment.}
\label{array}
\end{figure}

At TeV energies the flux of photons is low, the strongest steady source in
the sky is the CRAB nebula which gives only 0.4$\cdot $10$^{-11}$ photons/%
cm$^{2}$sec at 1 TeV. Large collection areas are required, of the order of a few$%
\cdot$ 10$^{4}$m$^{2}$, certainly impossible for space based experiments at
present. One therefore is confined to experiments on earth's surface. The
key feature of the presently most successful technique makes use of the air
as a Cherenkov medium, with changing index of refraction and nicely
transparent to the Cherenkov photons. The air is about 23 r.l. thick thus
confining the electromagnetic shower generated by the incoming TeV photon.
For stability of observational conditions clear air is needed preferentially
at an elevation of order 2500 m and above the cloud level. The Observatorio
del Roque de los Muchachos (ORM) on La Palma is such a location and has been
chosen by the HEGRA collaboration for the deployment of an extended
airshower array for TeV observations (see figure \ref{array}). Details of
the installations and properties of the detectors can be found in \cite{rho}%
. 
Of particular importance are the Air Cherenkov telescopes with large
detection area of about 30.000 $m^{2}$, good separation power of photon-
versus hadron showers and $\sim $ 1/$10^{\circ }$ angular resolution. Like
optical telescopes good observations are possible only in clear nights and
low moon light. The field of view of the telescopes is about $4^{\circ }$
and at any given setting essentially only one source can be observed.
Several telescopes combined (presently 4 of the 6 installed at ORM on La
Palma) give improved gamma hadron separation and also angular resolution at
the expense of some fraction of the detection area. The first source
definitely detected in the TeV-range has been the CRAB nebula~\cite{wee},
the remnant of the AD 1054 supernova that also houses the 33 msec CRAB
pulsar. As a mechanism for the generation of TeV photons in the nebula it is
assumed that electrons are accelerated to very high energies of order $10^{8}
$ GeV. The electrons produce synchrotron radiation in an ambient magnetic
field of the order of tens of $\mu $T. The synchrotron photon spectrum reaches
up to a few GeV and is detected by EGRET~\cite{ram}. Compton upscattering of
the synchrotron photons by the primary electrons is taken responsible for
the very high photons up to tens of TeV~\cite{gou2}. There must however
exist in the CRAB nebula an efficient mechanism to accelerate the electrons
to energies $\gg TeV$ with shock wave acceleration usually assumed as such a
mechanism. More sources of similar structure are to be expected in the
galaxy and indeed TeV photons from SN 1006 and from 1706-44 as similar
supernova remnants have been observed using the Air Cherenkov technique at a
site in Australia~\cite{kif}.

In a follow up observation of EGRET sources the Whipple collaboration
discovered TeV photons from one of the closest extragalactic sources in the
EGRET sample the Markarian galaxy Mkn 421 at a redshift of z=0.031 ($\sim $
300 Mill.ly.)~\cite{pun}. This observation was confirmed by HEGRA~\cite{pet}%
. In addition a similar galaxy, Mkn 501 at z=0.034 was detected as a rather
weak source and again confirmed by HEGRA~\cite{qui}. Both galaxies have an
active nucleus and belong to a subclass of AGN's, the so called Bl Lac
objects named after the first found galaxy of this type at a distance of z=0.069.
The active nucleus of Bl Lacs is at the center of an elliptical galaxy that
has only narrow emission lines of width $<5{\AA }$. In May 1996 the Whipple
telescope was lucky enough to detect two very big and extremely rapid flares
from Mkn 421 (missed by HEGRA because of daylight at La Palma)~\cite{gai}.
Most of the time both Mkn 421 and 501 are near the detection limit of the
telescopes however not inconsistent with the typical behavior of the flat
spectrum AGN's, that seem to change continuously their emission intensity
e.g. in the radio or optical region. 
\begin{figure}[tbp]
\begin{center}
\psfig{figure=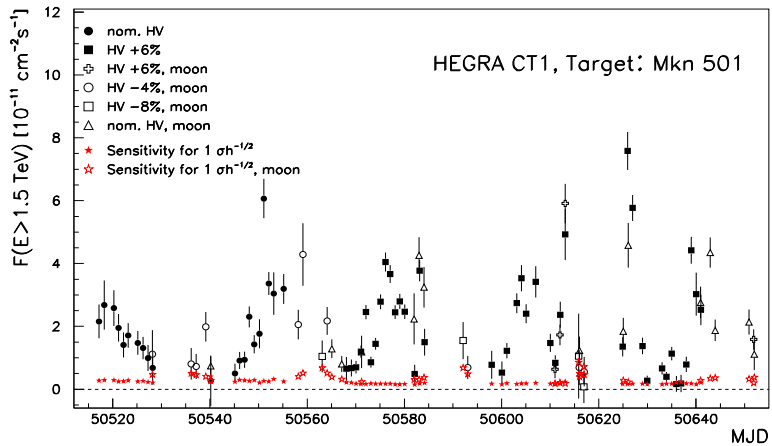,width=11.0cm,height=7.0cm}
\end{center}
\caption{The ``lightcurve'' of Mkn 501 from March to July 1997 at E $>$ 1.5
TeV. The data are preliminary. They have been obtained using the first
Cherenkov telescope CT1 of the HEGRA--collaboration.}
\label{lightcurve}
\end{figure}
When Mkn 501 became observable again this spring it was seen in a state of
rapid flaring (see figure \ref{lightcurve}) much brighter than 1996 and
frequently much more intense than the 'standard candle' CRAB, indeed up to
10 times the CRAB flux was observed making Mkn 501 the most intense TeV
gamma ray source in the sky. Furthermore the energy spectrum was observed to
extend well beyond 10 TeV with a seemingly unabsorbed power law with
spectral index $\sim $2.5 similar to the CRAB~\cite{aha2}. As an immediate
consequence, given the distance of Mkn 501, it became clear that the
universal infrared background must be very much lower than the upper limits
obtained so far and likely very close to the lower end of recent theoretical
estimates~\cite{fun,ste}. This opens up the possibility to really detect the
universal infrared background through observation of the absorption in the
spectra at some tens of TeV. The position of Mkn 501 can now be found with
TeV photons at the level of $0.01^{\circ }$, better than arcmin, which is
considered as a sort of entrance ticket to real astronomy. Mkn 501 is
observed to continue flaring at a high level and is meanwhile being detected
by several instruments using the Air Cherenkov technique~\cite{cat}.

\section{Terra Incognita at $>$100 TeV}

\label{Terra Incognita}

As shown in chapter \ref{The process gamma gamma} photons with energy $>$100
TeV get readily absorbed in the $2.7^{\circ }$K microwave background and as
a result the horizon out to which one can see into the universe comes down
to galactic distances. Only a faint halo of isotropic 'skyshine' is expected
to remain from photons created at $>$100 TeV~\cite{wdo}. However we get
knowledge of phenomena of energy at least six orders of magnitude higher
since airshowers up to $10^{20}$eV total energy (with a few events beyond)
have been observed. At very high energies $>$ $10^{19}$eV when charged
particles (say protons) would become stiff enough to keep direction in
intergalactic magnetic fields they may reveal sources in our local universe
out to about 50 Mpc, since on longer distances protons would be absorbed
through photon pion production in the $2.7^{\circ }$K photon background~\cite
{gre}. This possibility of 'proton astronomy' may soon be explored, when the
two 'Auger' arrays of size several thousand $km^{2}$ become operational. As
a further possibility we should be aware that proton acceleration to very
high energies $\gg $100 TeV in e.g. AGN should not only produce $\pi ^{\circ
}$ via photo pion production which may have -- as one of the possibilities --
been observed with 10 TeV photons from Mkn 421 and Mkn 501. It implies
charged pion production as well and therefore we obtain from $\pi ,\mu $
decay a muon- and electron neutrino flux of TeV energies. If detected it
surely tells of a similar photon flux, while the reverse is not necessarily
true since photons could have primary electrons as the only source.
Therefore it is of great importance to detect $>$100 TeV neutrinos from
distant sources to cover the highest energy window inaccessible to photon
detection. A rather promising project (AMANDA) to detect the very high
energy neutrinos is under construction using the ice of Antarctica at the
south pole station~\cite{ama}. Last antarctic summer several strings of
photomultipliers $\sim $400 m long have been successfully lowered to a depth
beyond 1500 m of ice. This setup should safely detect neutrinos originating
from cosmic ray interactions in earth atmosphere and it may come close to
detect (with some luck) extragalactic neutrinos. This would open up yet
another new window into the universe with entirely new information on the
most violent processes in the universe.

\section{Bursts of $\gamma$--rays}

\label{GRB}

\begin{figure}[tbp]
\begin{center}
\psfig{figure=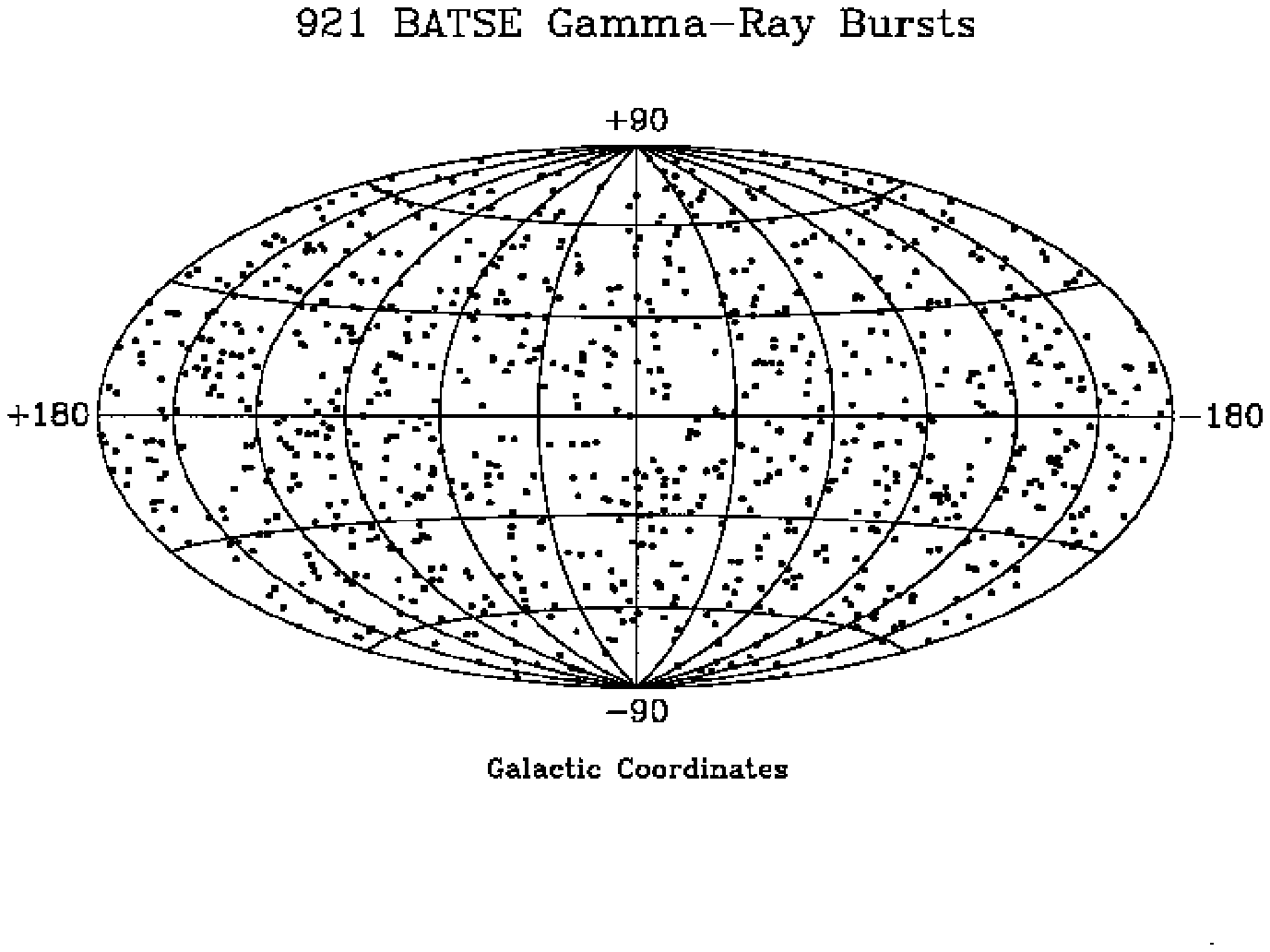,width=9.0cm}
\end{center}
\caption{Angular distribution of 921 GRBs in galactic coordinates.}
\label{bursts}
\end{figure}

An enigmatic astrophysical phenomenon, the so called $\gamma $--ray bursts
were serendipitously discovered in the early 70th in the course of
verification of the nuclear test ban treaty. Short bursts of $\gamma $--rays
in the keV-MeV range were observed ranging in time from msec to min. Nothing
else was seen at the $\gamma $--burst position that could reveal the nature
of phenomenon. A big surprise came with the CGRO which had with BATSE a
dedicated all sky monitor for $\gamma $--ray bursts on board. It detected
bursts at a rate of about one a day, more than 1500 by now. The distribution
on the sky was found to be completely isotropic (at the $<2\%$ level; see
figure \ref{bursts}) while the intensity distribution is inhomogeneous, less
frequent at low flux. The simplest hypothesis places them at cosmological
distances corresponding to a redshift z$\sim $1 which implies an energy
budget of $\sim 10^{48}$ erg/sec at the source. Recently a new attempt was
made to find counterparts with the deployment in space of the ``Beppo-Sax''
X--ray satellite. It allows for arcmin localization of $\gamma $--ray bursts
in the lower X--ray band within hours. A burst on 28/Feb./97 was the first
case where a fading counterpart was observed in the X-ray range by Beppo-Sax
as well as in the optical band using the William Herschel 4.2 meter
telescope at ORM on La Palma~\cite{gro}. Very recently on 8/May/97 again an
optical counterpart was detected at first getting brighter and than fading
again. This burst for the first time gave a lower limit on the distance of z$%
\geq $0.835 observing absorption lines features in the spectrum using the
KEK telescope on Mauna Kea, Hawaii. More such observations might finally
reveal an underlying mechanism, the present observations being fully
consistent with a nonisotropic relativistically expanding plasma as the
source of $\gamma $--ray bursts~\cite{mes}. This resembles similarity with
radio blobs in AGN jets however many more accurate pointings and counterpart
detections are needed to find a clue.

\section{Closing Remark}

\label{Closing Remark} This century has seen a dramatic extension of
knowledge about the universe in particular exploring observational
possibilities in e.g. the radioband, the X--ray range and very recently up
to the TeV photons. Entirely new information should come from very high
energy neutrino observation. A further leap into new territory is sure to
come once gravitational waves will have been detected. Exciting times at new
frontiers in astronomy are ahead of us.

\section*{Acknowledgments}
I would like to thank my colleagues in HEGRA for numerous  fruitful
discussions on the subject of this talk. I thank the organizers of
Photon 97 for the opportunity to present this matter at the meeting.
Finally I am indebted to U. Kleinevoss for help with the manuscript.

\end{document}